\documentclass[11pt,a4paper]{article}

\usepackage{a4wide}
\usepackage{xspace}
\usepackage{multicol}
\usepackage{amsmath}
\usepackage{graphicx}
\usepackage{epsfig}
\usepackage{hyperref}
\usepackage{url}
\usepackage{amssymb}

\newcommand{\ie}{\emph{i.e.}\xspace}
\newcommand{\eg}{\emph{e.g.}\xspace}

\def\order#1{{\cal{O}}\left(#1\right)}
\newcommand{\avg}[1]{\left\langle #1 \right\rangle}

\newcommand{\NLO}{\mathrm{NLO}}
\newcommand{\GeV}{\,\mathrm{GeV}}
\newcommand{\TeV}{\,\mathrm{TeV}}

\def\ratio{{\cal R}}

\title{\textbf{A simple description of jet cross-section ratios}}
\author{Gregory Soyez\\
\\
{\sl \small Institut de Physique Th\'eorique, CEA Saclay, CNRS URA 2306, F-91191 Gif-sur-Yvette, France}}
\date{}

\begin{document}

\maketitle

\begin{abstract}
  We compute the ratio of the inclusive jet cross-sections obtained
  with the same jet algorithm at two different values of the jet
  radius. We perform a computation of that observable at NLO
  ($\order{\alpha_s^2}$) in perturbative QCD and compute
  non-perturbative corrections from soft-gluon emission. We discuss
  predictions for RHIC and the LHC.
\end{abstract}

\paragraph{Introduction}\label{sec:intro}

In this letter, we are interested in computing the ratio of the
inclusive jet cross-section computed with the same jet algorithm at
two different values of $R$:
\begin{equation}\label{eq:def}
\ratio(p_t;R_1,R_2) = \frac{\frac{d\sigma}{dp_t}(R=R_1)}{\frac{d\sigma}{dp_t}(R=R_2)}.
\end{equation}
Our main objective is to show that the minimal effort required to get
a reliable prediction for $\ratio$ is to include $\order{\alpha_s^2}$
perturbative correction as well as (universal) non-perturbative power
corrections.

We shall first discuss the perturbative computation of $\ratio$. This
is interesting {\em per se} since, as we will see below, instead of
computing naively the ratio of the cross-sections computed at NLO,
which would formally correspond to a computation of $\ratio$ up to
$\order{\alpha_s}$, it is actually possible to obtain the
$\order{\alpha_s^2}$ corrections \cite{dms}\footnote{See also
  \cite{zeus} for an experimental measurement and QCD computation of
  jet cross-section ratios with different jet algorithms and a fixed
  $R$.}.

Since jets are basic observables at the LHC, and both ATLAS and CMS
plan to use the anti-$k_t$ algorithm \cite{antikt} with two different
radii ($R=0.4$ and $0.6$ for ATLAS, $R=0.5$ and $0.7$ for CMS), they
could in principle measure the ratio. Compared to the measurement of
the inclusive jet cross-section, the ratio would not have the
uncertainty on the luminosity measurement and would probably be less
sensitive to the jet energy scale.
Below, we shall compare our NLO QCD predictions with and without
hadronisation corrections to the ATLAS recent measurements
\cite{atlas_jets} and make predictions for the cross-section ratio. 

Another situation in which the cross-section ratio is an interesting
observable is at RHIC where it can be measured both in
proton-proton and heavy-ion collisions\footnote{Similar considerations
  would hold for PbPb collisions at the LHC, with the extra
  complication that the energy of the collision differs from the $pp$
  one.}. Due to the interaction with the hot medium produced in
heavy-ion collisions, one expects the jet to loose energy and to be
broadened. That would directly translate into a decrease of the
cross-section ratio (see \eg \cite{vitev} for a computation with and
without medium effects at $\order{\alpha_s}$ in QCD). Here we shall
show that, for the $pp$ reference measurement, the next order and
hadronisation bring large corrections.

\paragraph{Perturbative expansion}\label{sec:pert}

Let us start with a perturbative QCD computation of the cross-section
ratio. Naively, since inclusive jet cross-sections are known up to NLO
accuracy ($\order{\alpha_s^3}$), one would use\footnote{For
  readability, we use $\sigma(p_t;R)$ as a shorthand notation for
  the differential cross-section $\frac{d\sigma}{dp_t}(R)$.}
\begin{equation}\label{eq:naive}
\ratio(p_t;R_1,R_2) = \frac{\sigma^{\NLO}(p_t;R_1)}{\sigma^{\NLO}(p_t;R_2)},
\end{equation}
which is formally an $\order{\alpha_s}$ computation of $\ratio$.

The interesting point is that, by making the perturbative expansion
explicit, $\ratio$ can actually be computed up to
$\order{\alpha_s^2}$. To see this, consider the perturbative expansion
of the jet cross-section:
\[
\sigma(p_t; R)=\alpha_s^2 \sigma^{(2)}(p_t) + \alpha_s^{(3)} \sigma^{3}(p_t;R) + \alpha_s^4 \sigma^{(4)}(p_t;R) + \order{\alpha_s^5},
\]
where we have taken into account the fact that the leading-order
contribution does not depend on $R$. The contribution
$\sigma^{(n)}(p_t;R)$ at a given order $\alpha_s^n$ receives
contributions from tree diagrams with $n$ particles in the
final-state, up to $(n-2)$-loop diagrams with 2 particles in the final
state. Denoting by $\sigma^{(n,p)}(p_t;R)$ the $p$-loop contribution
to $\sigma^{(n)}(p_t;R)$, we have
\begin{eqnarray}
\sigma(p_t; R)
& = & \alpha_s^2 \sigma^{(2,0)}(p_t)
  +   \alpha_s^3 [\sigma^{(3,0)}(p_t;R) + \sigma^{(3,1)}(p_t)]\nonumber\\
& + & \alpha_s^4 [\sigma^{(4,0)}(p_t;R) + \sigma^{(4,1)}(p_t;R) + \sigma^{(4,2)}(p_t)]
  +   \order{\alpha_s^5},\label{eq:xsect_exp}
\end{eqnarray}
where we have again used the fact that the contributions with only 2
particles in the final state do not depend on the jet radius $R$.

If we use (\ref{eq:xsect_exp}) to expand the ratio $\ratio(R_1,R_2)$ in series
of $\alpha_s$, we obtain, up to corrections of order $\alpha_s^3$,
\begin{eqnarray}\label{eq:ratio_pqcd}
\ratio(p_t;R_1,R_2)
& = &1
 + \alpha_s   \frac{\Delta\sigma^{(3,0)}(p_t;R_1,R_2)}{\sigma^{(2)}(p_t)}\\
& + &\alpha_s^2 \frac{\Delta\sigma^{(4,0)}(p_t;R_1,R_2) + \Delta\sigma^{(3,1)}(p_t;R_1,R_2)}{\sigma^{(2)}(p_t)}
 - \alpha_s^2 \frac{\sigma^{(3)}(p_t;R_2) \Delta\sigma^{(3,0)}(p_t;R_1,R_2)}{[\sigma^{(2)}(p_t)]^2},\nonumber
\end{eqnarray}
with $\Delta\sigma^{(n,p)}(p_t;R_1,R_2)=\sigma^{(n,p)}(p_t;R_1)-\sigma^{(n,p)}(p_t;R_2)$.

The remarkable fact, that allows for the computation to be performed
at $\order{\alpha_s^2}$, is that the two-loop contribution to the NNLO
jet-cross-section, that prevents one from obtaining a NNLO computation
of the inclusive jet cross-section (\eg using NLOJet~\cite{nlojet}),
does not appear in the computation of the cross-section
ratio\footnote{Note that it would contribute at the next order.}.

In what follows, the LO ratio will refer to (\ref{eq:ratio_pqcd}) with
the two first terms kept --- the $\order{\alpha_s}$ expansion, \ie the
first non-trivial order ---, while the NLO ratio will also incorporate
the $\order{\alpha_s^2}$ corrections in (\ref{eq:ratio_pqcd}). 

Before proceeding with the discussion about non-perturbative effects,
it is interesting to comment a bit on eq. (\ref{eq:ratio_pqcd}). In
the collinear limit, the NLO (resp. NNLO) correction to the
cross-section will be proportional to $\alpha_s\log(1/R)$
(resp. $\alpha_s^2 \log^2(1/R)$), which would be the dominant
correction at small jet radius. In the computation of the ratio, the
$\order{\alpha_s}$ term only involves the cross-section difference and
will thus be proportional to $\log(R_1/R_2)$ while the next order will
involve $\log(R_1/R_2)\log(1/R_2)$. This means that for $R_1\sim
R_2\ll 1$, the collinear contribution will mostly appear from NLO
onwards and we may thus expect large NLO corrections.

\paragraph{Non-perturbative corrections}\label{sec:hadr}

As we shall see later when making explicit computations of the ratio
$\ratio$, for small values of $R$, hadronisation corrections may have
a significant impact on the jet cross-section and thus on
$\ratio$. One could in principle rely on Pythia \cite{pythia} or
Herwig \cite{herwig} (or, better, a combination of both) in order to
estimate the correction factor one has to apply to go from a
parton-level cross-section to a hadron-level cross-section, \ie to
estimate hadronisation corrections.
Keeping in mind that we want to provide as simple a description of the
cross-section ratio as we can, we shall instead give an analytic
estimate of the hadronisation corrections. In \cite{dms},
hadronisation correction are computed from soft-gluon emission and the
authors obtain that the effect of hadronisation is to shift the $p_t$
of the jet by an average amount
\begin{equation}\label{eq:hadr_shift}
\avg{\delta p_t}_{\rm hadr} = \frac{-2C_R}{R} \frac{2M}{\pi} {\cal A}(\mu_I).
\end{equation}
In that expression, $C_R$ if the Casimir factor which should be $C_F$
for quark jets and $C_A$ for gluon jets, $M$ is the Milan factor that
depends on the jet algorithm --- it is universal
\cite{universalmilan}, $M\approx 1.49$, for the anti-$k_t$ algorithm
while, for the $k_t$ algorithm, one finds \cite{ktmilan} $M\approx
1.01$ ---, and ${\cal A}(\mu_I)$ carries all the non-perturbative
dependence. The latter can be rewritten\footnote{At the 2-loop
  accuracy and in the $\overline{{\rm MS}}$ scheme.} as
\begin{equation}\label{eq:amui}
{\cal A}(\mu_I) = \frac{\mu_I}{\pi}\left[
  \alpha_0(\mu_I) - \alpha_s(p_t) 
  - \frac{\beta_0}{2\pi}
    \left(\log\left(\frac{p_t}{\mu_I}\right)+\frac{K}{\beta_0}+1\right)
    \alpha_s^2(p_t)
\right],
\end{equation}
where the average coupling in the infrared region
$\alpha_0(\mu_I)=(1/\mu_I)\int_0^{\mu_I}\alpha_s(k_t)dk_t$ is
frequently encountered in event-shape studies (see \eg
\cite{eventshapes}), $\beta_0=(11 C_A-2 n_f)/3$ and
$K=C_A\left(\frac{67}{16}-\frac{\pi^2}{6}\right)-\frac{5}{9}n_f$.

Including the hadronisation corrections to the perturbative cross-section can then
be done using\footnote{In practice, since quark and gluon-jets have a
  different $p_t$ shift due to hadronisation, one should consider
  their contributions separately.}
\begin{equation}\label{eq:hadronisation}
 K_{\rm hadr}(p_t; R)
 = \frac{\sigma(p_t;R)}{\sigma_{\rm pQCD}(p_t; R)}
 \approx \frac{\sigma_{\rm pQCD}(p_t-\avg{\delta p_t}_{\rm hadr};R)}{\sigma_{\rm pQCD}(p_t; R)}
 \approx \frac{\sigma_{\rm LO}(p_t-\avg{\delta p_t}_{\rm hadr};R)}{\sigma_{\rm LO}(p_t; R)}.
\end{equation}
For the first equality, we have neglected the dispersion in $\delta
p_t$ (\ie assumed that the shift was always the average one) which
would correspond to higher power corrections that are not as well
controlled from LEP data. Approximating the full perturbative
cross-section by the leading-order expression in the second equality
is motivated by the fact that the computation of hadronisation
corrections from soft-gluon emission is done for the underlying $2\to
2$ scattering \ie from the leading-order process.

Finally, the cross-section ratio after taking into account the
hadronisation corrections is
\begin{equation}\label{eq:ratio_hadron}
\ratio(p_t;R_1,R_2) = 
\frac{K_{\rm hadr}(p_t; R_1)}{K_{\rm hadr}(p_t;R_2)}
\ratio_{\rm pQCD}(p_t;R_1,R_2),
\end{equation}
with $\ratio_{\rm pQCD}(p_t;R_1,R_2)$ computed from
eq. (\ref{eq:ratio_pqcd}).

Because of the $1/R$ behaviour of (\ref{eq:hadr_shift}), we may also
expect sizeable effects from the non-perturbative corrections at small
$R$. Note however that the factor of the $1/R$ term is rather small
($2C_F{\cal A}(\mu_I)\approx 0.5\GeV$), compared to the corresponding
QCD corrections that would typically scale like $\alpha_s p_t$ and so
dominate at moderate $R$ and $p_t$.

\paragraph{Comparison with experiments}\label{sec:exp}

In the following lines, we briefly discuss the perturbative
computation of $\ratio$ and the hadronisation corrections at two
different energies: RHIC ($\sqrt{s}=200 \GeV$) and the LHC
($\sqrt{s}=7 \TeV$).

As far as the perturbative part of the computation is concerned, we
have used NLOJet (v4.1.2) \cite{nlojet} for the computation of the
different pieces in (\ref{eq:ratio_pqcd}). We have considered the
CTEQ6.6 NLO PDF set \cite{cteq66} as well as the MSTW08 NLO and NNLO
sets \cite{mstw08} though, for brevity, we shall only show the CTEQ6.6
results in what follows. The scale uncertainties have been
obtained\footnote{For both scales we compute a negative and a positive
  uncertainty. The renormalisation and factorisation scale
  uncertainties are then added in quadrature to obtain the total
  uncertainty.} by varying independently the renormalisation and
factorisation scales from $p_{t,{\rm jet}}$ to $p_{t,{\rm jet}}/2$ and
$2p_{t,{\rm jet}}$.

\begin{figure}
\includegraphics[angle=270,width=0.5\textwidth]{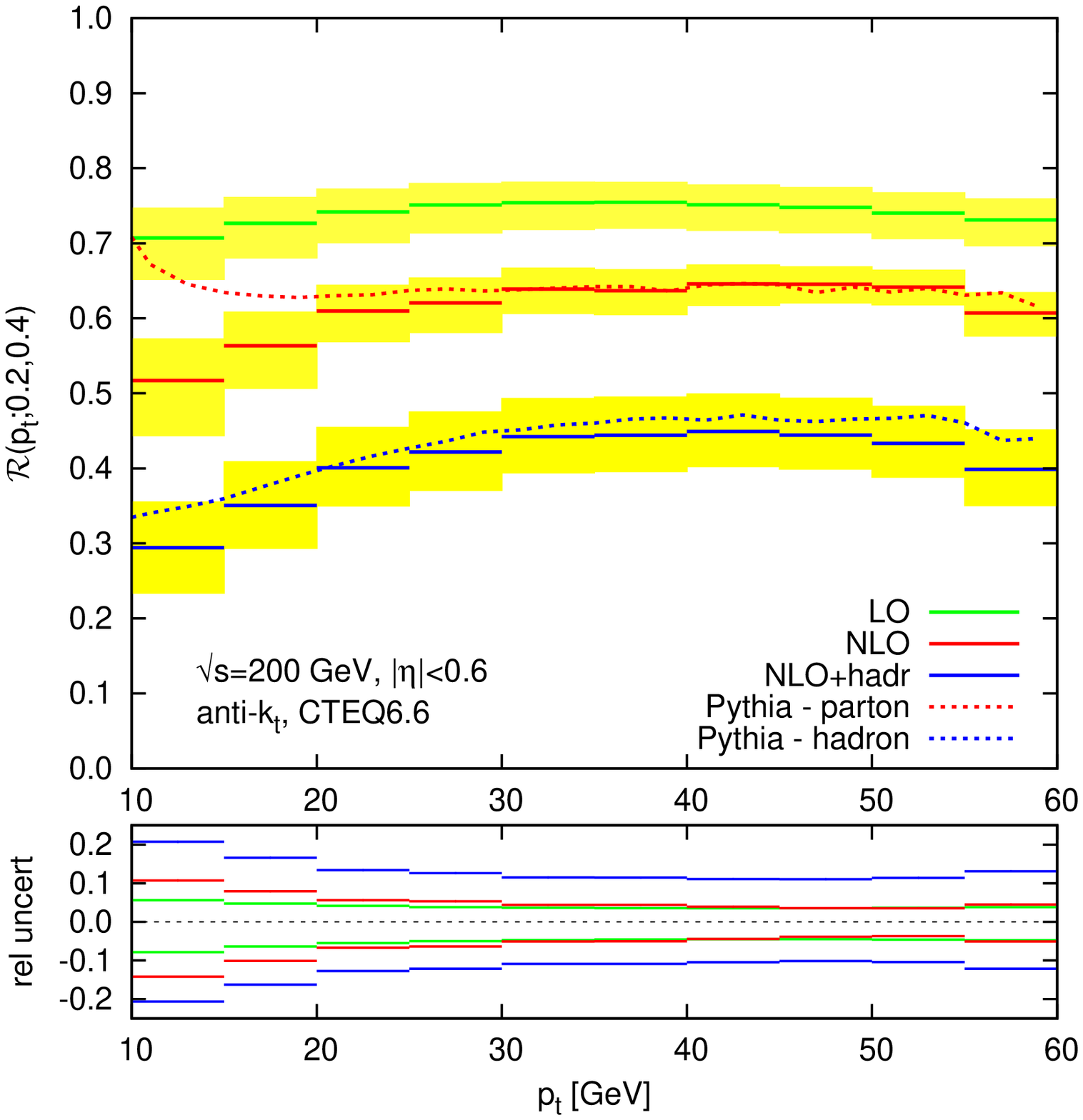}
\includegraphics[angle=270,width=0.5\textwidth]{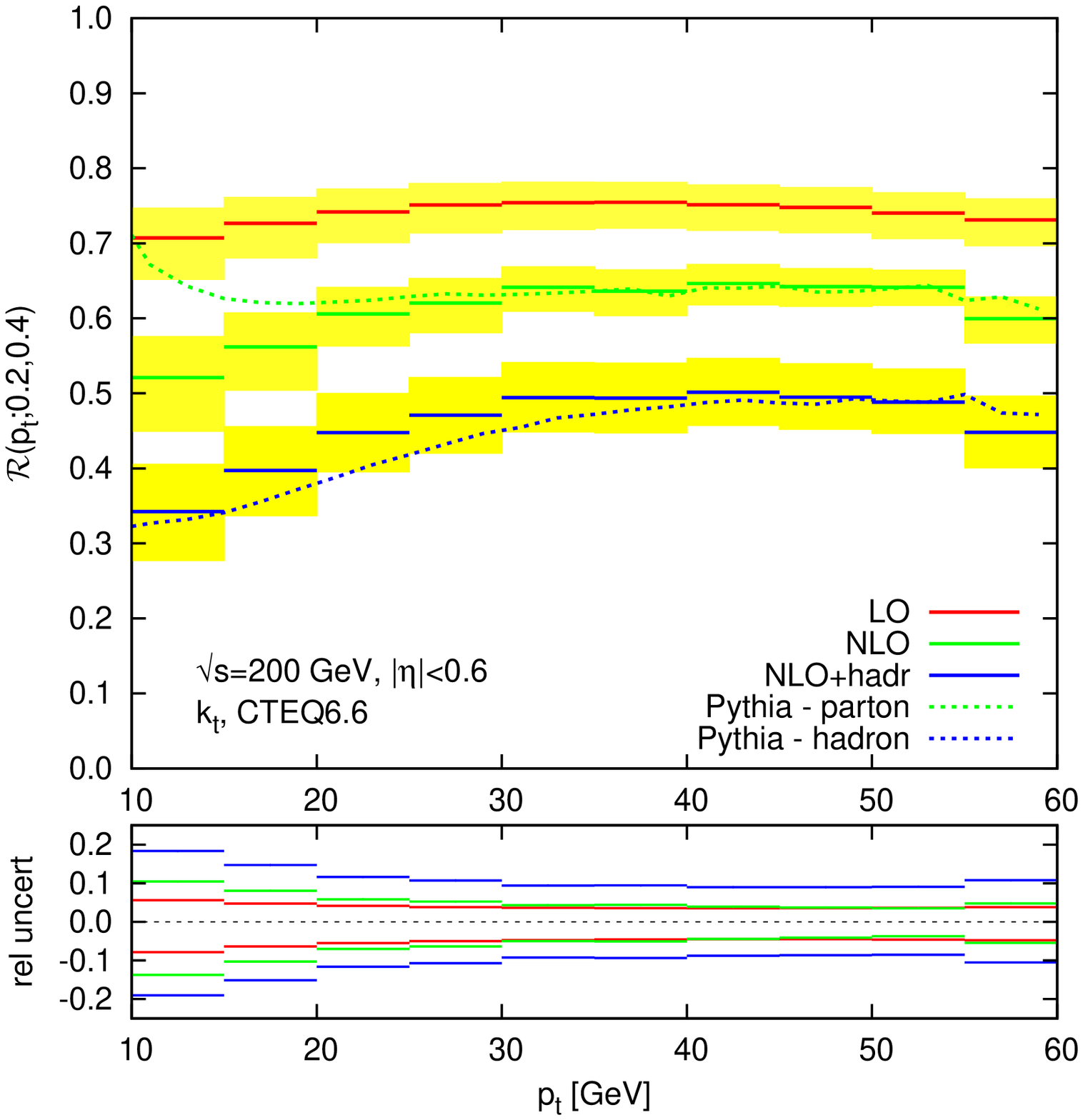}
\caption{QCD predictions for the ratio $\ratio(p_t;0.2,0.4)$ at RHIC
  ($\sqrt{s}=200\GeV$) for the anti-$k_t$ (left) and $k_t$ (right) jet
  algorithms. On the top panel, the solid lines correspond, from top
  to bottom, to the LO QCD computation (green), to the NLO QCD ratio
  (red) and to the NLO QCD computation including hadronisation
  effects (blue). The uncertainties due to the scale choice and, when
  relevant, hadronisation are shown as shaded bands on the top panel
  and the relative scale uncertainty is plotted on the bottom
  panel. For comparison, we have also plotted in dashed lines the
  parton-level (red) and hadron-level (blue) predictions from
  Pythia.}\label{fig:star}
\end{figure}

To compute the hadronisation corrections, the only parameter we
need\footnote{We will always consider large-enough $p_t$ so we can
  safely use $n_f=5$ in (\ref{eq:amui}) and, for consistency, we have
  used the running coupling provided together with the PDF set.}  is
$\alpha_0$. As already mentioned, this can be extracted
\cite{eventshapes} from event-shape distributions at LEP and we shall
use the value $\alpha_0(\mu_I)=0.503$ with $\mu_I=2 \GeV$, obtained
from JADE data \cite{alpha0jade}. The uncertainty on the hadronisation
corrections will be estimated by varying the Milan factor ($M=1.49$
for the anti-$k_t$ algorithm and $M=1.01$ for $k_t$) by the standard
20\%.

Let us start by discussing the case of RHIC, where STAR is planning to
measure \cite{star} the ratio $\ratio(p_t; 0.2, 0.4)$ for both the
$k_t$ \cite{kt} and anti-$k_t$ \cite{antikt} algorithms. Though the
ratio will be measured in proton-proton and gold-gold collisions with
the ultimate goal to see jet-broadening effects due to interaction
with the hot medium produced in heavy-ion collisions, we just focus on
the $pp$ case here\footnote{See \cite{vitev} for a LO description of
  $\ratio$ for $pp$ and gold-gold collisions, incorporating medium
  effects for the latter.}.
The result is presented in Fig. \ref{fig:star} for both
algorithms. The first message is that NLO corrections to $\ratio$ are
substantial ($\sim 0.1-0.15$) and, probably as a consequence, the
scale uncertainty does not decreases when going from LO to NLO. Though
they are strictly the same only at LO, the $k_t$ and anti-$k_t$
algorithms show a very similar cross-section ratio also at NLO.
Then, as a consequence of the choice of rather small values of $R$,
hadronisation effects are also sizeable ($\sim 0.15-0.2$). In this
case, since the Milan factor is a bit larger for the anti-$k_t$
algorithm than for $k_t$, the final ratio tends to be a bit larger for
the $k_t$ algorithm.
Finally, Fig. \ref{fig:star} shows that the NLO pQCD computation of
$\cal{R}$ is in good agreement with what is obtained from
Pythia\footnote{For Pythia simulations, the ratio is obtained by
  explicitly dividing the jet cross-section computed with the two
  radii.} (v6.4) at parton-level (\ie including parton shower from
initial and final-state radiation), and our final prediction,
including non-perturbative corrections is also in good agreement with
what Pythia predicts when hadronisation is included\footnote{The
  underlying-event corrections could also be taken into account both
  in our computation and in the Pythia simulation but they have a very
  small impact on $\ratio$.}.

\begin{figure}
\includegraphics[angle=270,width=0.5\textwidth]{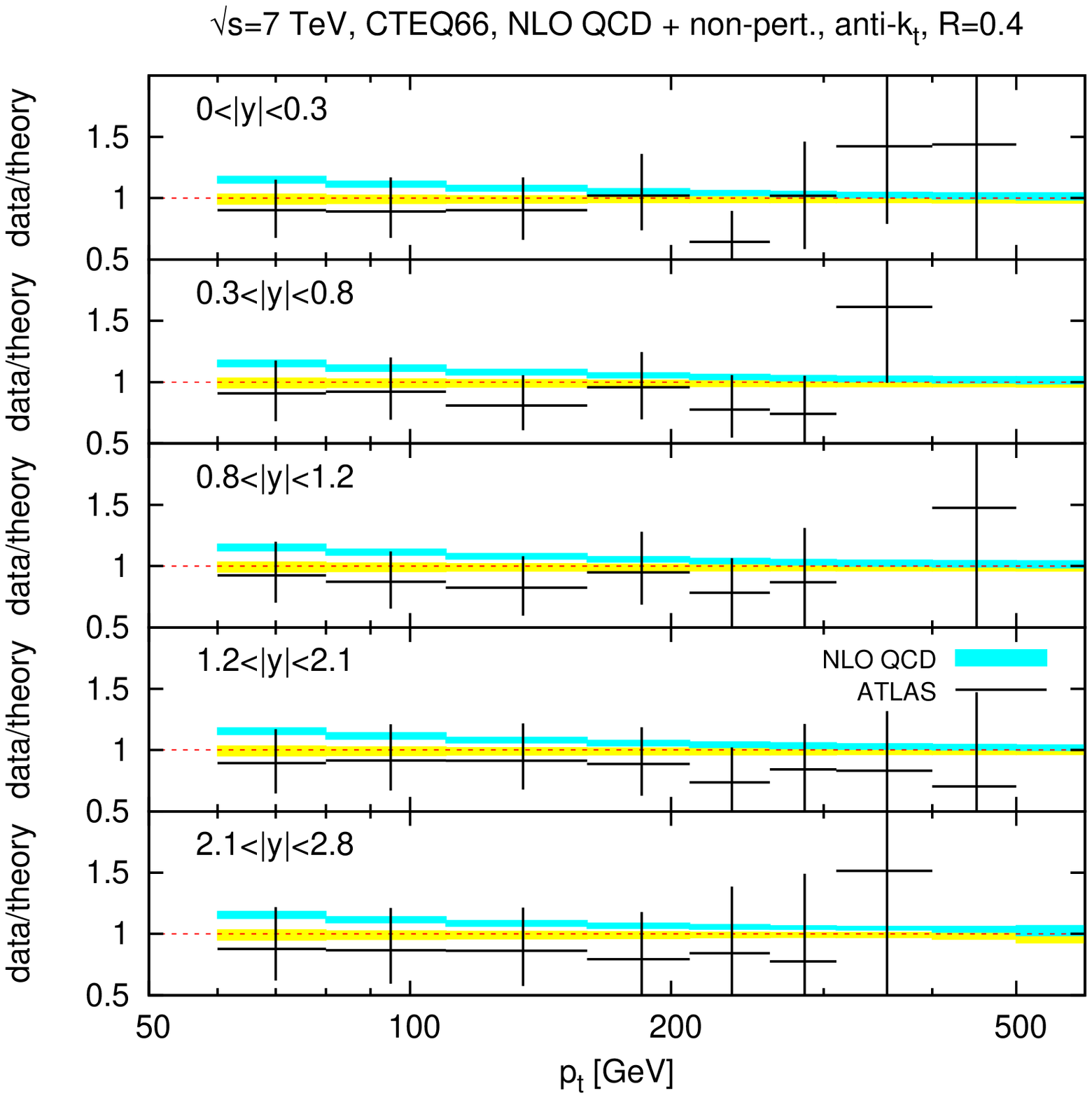}
\includegraphics[angle=270,width=0.5\textwidth]{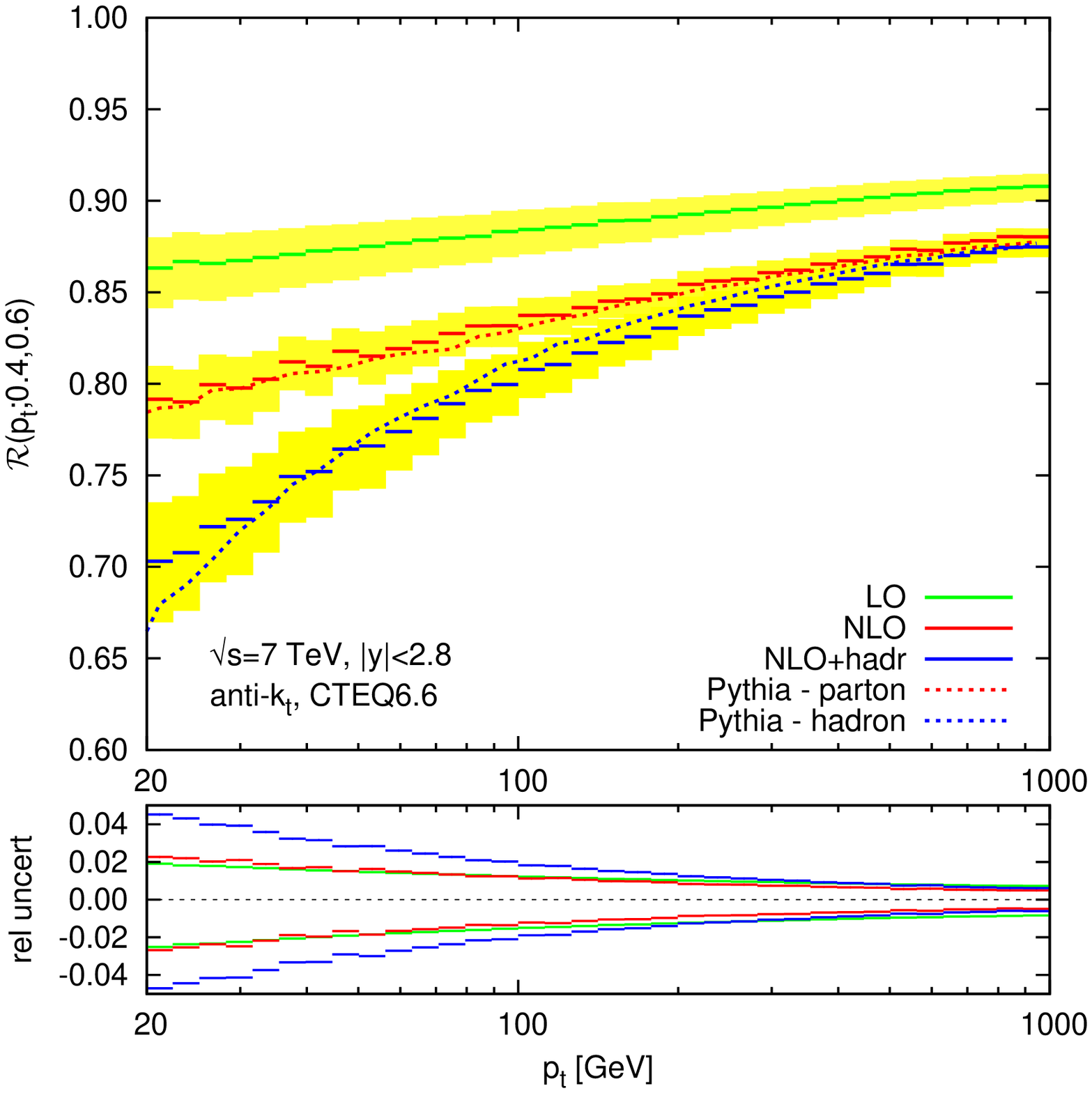}
\caption{Left: comparison of our inclusive jet cross-section
  computations with the ATLAS measurements. The ratio between the
  experimental values and the theory predictions (NLO QCD including
  non-perturbative effects) is plotted; the yellow band represents the
  uncertainty on the theoretical prediction and the cyan band
  corresponds to the NLO QCD prediction without hadronisation
  corrections. Right: predictions for the ratio $\ratio(p_t;0.4,0.6)$
  for the anti-$kt$ algorithm at ,from top to bottom, LO, NLO, NLO
  with hadronisation corrections. See Fig. \ref{fig:star} for
  conventions.}\label{fig:atlas}
\end{figure}

We now turn to the case of measurements at the LHC and, more
precisely, to the jet cross-section measured very recently at
$\sqrt{s}=7\TeV$ by the ATLAS collaboration \cite{atlas_jets}. On
Fig. \ref{fig:atlas} we have plotted our predictions both for the jet
cross-section (anti-$k_t$ algorithm with $R=0.4$) and the
cross-section ratio $\ratio(p_t; R_1=0.4,R_2=0.6)$. The jet
cross-section is compared to the ATLAS measurements and we see that,
though the pure NLO QCD prediction (cyan band) describes the data
nicely, the inclusion of the non-perturbative power corrections
improves the description. Note also that the non-perturbative
corrections obtained in our approach are compatible with the numbers
obtained from Pythia and Herwig and quoted by ATLAS. If we now
consider the cross-section ratio, see the right plot on
Fig. \ref{fig:atlas}, we basically recover the main features already
discussed in the case of RHIC. However, both the NLO QCD corrections
and the hadronisation corrections are reduced compared to what we
observed at RHIC. This is even more true for the non-perturbative
corrections at large $p_t$ which become very small. This is likely due
to two effects: first, the considered radii are larger, reducing the
effect of the collinear divergence in the NLO QCD computation as well
as the hadronisation corrections that behave like $1/R$. Then, the
inclusive jet cross-section is much less steep at the LHC than at RHIC
and thus a common $p_t$ shift would have a larger impact at RHIC.

\paragraph{Conclusions}\label{sec:ccl}

To summarise, we have discussed in this letter the minimal ingredients
needed to get a reliable calculation of the ratio
$\ratio(p_t;R_1,R_2)$ of the $p_t$-dependent inclusive jet
cross-section computed with the same jet algorithm at two different
values, $R_1$ and $R_2$, of the jet radius.

We have seen that by making an explicitly expansion in powers of
$\alpha_s$, we can compute $\ratio$ perturbatively at
$\order{\alpha_s^2}$, the NLO accuracy for that observable, that is
one order higher than what we would naively expect from the direct
ratio of the cross-sections.  The explicit computation of $\ratio$ at
NLO can be done {\em e.g.}  using the NLOJet++ event generator. Note
that using techniques of \cite{loopsim} would allow us to obtain an
approximate NNLO calculation and further test the convergence of the
perturbative series.

Then, we have estimated the non-perturbative corrections to the
ratio. They are based on universal power corrections and the only free
parameter, $\alpha_0$, can be estimated from fits to event-shape
measurements at LEP. 

Finally, we have seen that, in practice, both the $\order{\alpha_s^2}$
terms and the non-perturbative effects are numerically sizeable,
except for the hadronisation correction at large $p_t$. In the case of
the recent jet measurements done by ATLAS, it would be interesting to
see if the computation of the ratio could benefit from reduced
uncertainties compared to the jet cross-section itself.

In the case of RHIC, the NLO and hadronisation corrections are even
larger. It is important to keep that in mind when performing the same
computation for heavy-ion collisions, in the presence of the medium:
the one-gluon-emission approximation is likely to be insufficient. One
has to include the next-order corrections as well as non-perturbative
effects.

\paragraph{Acknowledgements}  
I am grateful to Gavin Salam and Lorenzo Magnea for useful information
about the theory and phenomenology of hadronisation corrections. I
would also like to thank Gavin Salam and Matteo Cacciari for a careful
reading of the manuscript. This work was supported in part by grant
PITN-GA-2010-264564.


\end{document}